# An Efficient Key Management Scheme For In-Vehicle Network

Hsinlin Tan

*Abstract*—Vehicle technology has development rapidly these years, however the security measures for in-vehicle network does not keep up with the trend. Controller area network(CAN) is the mostly used protocol in in-vehicle network. With the characteristic of CAN, there exists many vulnerabilities including lacks of integrity and confidentiality, and hence CAN is vulnerable to various attacks such as impersonation attack, replay attack, and etc. In order to implement the authentication and encryption, secret key derivation is necessary. In this work, we proposed an efficient key management scheme for in-vehicle network. In particular, the scheme has five phases. In the first and second phase, we utilize elliptic curve cryptography based key encapsulation mechanism(KEM) in order to derive pair-wise secret between each ECU and a central secure ECU in the same group. Then in the third phase, we design secure communication to derive group shared secret among all ECU in a group. In the last two phases, SECU is not needed, regular ECU can derive session key on their own. We presented a possible attack analysis(chosen-cipertext attack as main threat) and a security property analysis for our scheme. Our proposed scheme is evaluated based on hardware-based experiment of three different microcontrollers and software-based simulation of IVNS. We argue that based on our estimation and the experiment result, our scheme performs better in communication and computation overhead compared to similar works.

## I. INTRODUCTION

With the swift development of artificial intelligence and vehicular technology, autonomous vehicles services like self-driving taxi are becoming more and more realistic. According to Fortune Business Insights, the global autonomous cars market is projected to grow from $1.64 billion in 2021 to $11.03 billion in 2028 at a CAGR of 31.3% in the 2021-2028 period[1]. Giant businesses associated with automobile manufacturing are putting more interest in autonomous vehicles. Recently, self-driving tech company Pony.ai, backed by Toyota Motor Corporation and Tencent, has obtained a taxi license in China, which will allow them to operate 100 robotaxis in the Guangzhou city district of Nansha[2].

However, many vulnerabilities exist in automobile. There have been numerous attacks against the in-vehicle network. Various examples showing that vehicle from different motor corporation can be hacked at certain situation. In [3], researchers from Tencent Keen Lab discovered security vulnerability on Tesla Model X and implemented remote control of the vehicle. Tesla reminded their customer to update their firmware as the response of experiment. In [4], Charlie and Chris successfully performed remote attack against an unaltered 2014 Jeep Cherokee and similar vehicles that results in physical control of some aspects of the vehicle, including steering, braking, and etc. In their work, a list of the potential entry points for an attacker is given, which consists of Bluetooth, remote keyless entry, and etc. Arguably, any piece of technology that interacts with the outside world could be a potential entry point for attacker. As a result of the experiment, Chrysler voluntarily recalls 1.4 million vehicles. In [5], Charlie and Chris took one step further from their previous work. Last time they can only disable the car's brakes at low speeds, however, now they are able to bring the vehicle to a halt from any speed in seconds. They can also digitally turn the wheel themselves at any speed.

Generally, each individual electronic control unit (ECU) is responsible for monitoring and controlling one vehicle feature. In a modern vehicle, the number of ECU can be ranged from 50 to 100, with each ECU responsible for one or more functions. Among various bus protocol used in vehicle for data transmission, Controller Area Network(CAN) protocol is the most widely used to sustain critical functions. However, with the characteristics of broadcast communication, no authentication or encryption, and arbitration scheme, CAN bus does not satisfy security requirements such as integrity, confidentiality, authentication and etc, which may lead to various attacks.

Many researchers have been committed to implementing and testing security of CAN bus traffic. Most of the work can be categorized into two groups based of the mechanism they use, one is in-vehicle intrusion detection system(IDS), and the other one is cryptography. Using IDS to detect malicious message is an important approach for CAN bus security. Various methods have been proposed by many researchers recently[6]-[9]. However, there are still many downsides of IDS. For example, machine learning based IDS may have high computational resource requirements, yet ECU resources that exist inside vehicles may not be able to handle this workload. As for IDS similar to[27], researchers install IDS into every ECU for anomaly detection, this will eliminate the high cost of computation. However, incompatibility and high deployment cost will be their new problems. We will not further discuss details on IDS, since it is out of the scope of this work. Various cryptography approaches were proposed to solve the problems brought by the vulnerabilities of CAN bus[10-24], [30], and [34]. Cryptography mechanism can be used in CAN bus to provide authentication, data integrity with Message Authentication Code(MAC), and confidentiality together with symmetric and asymmetric cryptosystems. The main challenges of implementing cryptography schemes in CAN bus are adding extra payload size on the bus and total cryptosystems execution time. These cryptography approaches will be explained more on Section II. Most of the research are focusing on building cryptosystems for message transmission, in this scenario, pre-shared symmetric key among ECU on CAN bus is necessary. However, not much attention have been paid on the derivation of pre-shared symmetric key.

In this work, we proposed an key management scheme on in-vehicle network. There are five phases of our scheme in total. First part is setting up security parameters. In the second part, exchanging symmetric key between an ECU and the secure ECU using the Diffie-Hellman key encapsulation mechanism. In the third part, the group shared secret is exchange among every ECU and the master ECU using the symmetric key established in second part. Then in the fourth part, any ECU on the group can be responsible for sending a random seed to other ECU to generate the group shared session key. In the last part of our scheme, when the counter on ECU reach its limit, each ECU can update their session key in real time without any message being transmission. To the best of our knowledge, we argue that our work is the first one to employ asymmetric cryptography(KEM) in only one phase in a key management scheme without the long-term pre-shared key.

Our main contributions are summarized as follows:
1) We proposed a novel efficient key management scheme for in-vehicle network without the long-term pre-shared key. The scheme is based on KEM together with symmetric cryptography and hash function plus a simple yet effective session key refreshing method. The scheme we proposed has the advantage of lower communication and computation overhead compared to similar works.
2) We provide a formal security analysis for our KEM related to the chosen-ciphertext attacks. Then security property analysis(cryptoperiod & security strength) is presented based on NIST 800-57. Lastly, other possible attack scenarios that our scheme may encounter are discussed.
3) We evaluate the performance of our scheme based on two experiment. Hardware-based is implemented using three different micro-controllers(Arduino UNO R3, WinnerMicro W806 and STM32F407ZG) and software-based simulation is built via in-vehicle network simulator. Our scheme is superior than similar works in terms of communication overhead based on the experiment result.

The rest of the work is organized as follows. In section II, we summarize and discussed most recent work of in-vehicle network security. In section III, we introduced the concept of key encapsulation mechanism and the system model of our work. In section IV, we presented and explained the key management scheme on CAN bus we proposed, and discussed the security problems of the scheme. In section V, we presented the implementation of our scheme, and the performance of our scheme is presented and compared with other approaches. In section VI we make the conclusion.

## II. RELATED WORK

In this section, we summarized and discussed previous work on exploiting in-vehicle network vulnerability and security approach on in-vehicle network, in which most in-vehicle network are CAN and CAN-FD. Table I presented the summary of most research on in-vehicle security, namely, the features, implementation platform and shortcomings of each work are shown.

### A. Research on the Vulnerabilities of In-Vehicle Networks

Various researches have been conduct on exploiting the vulnerabilities of in-vehicle networks[4], [21], [38], and [40]. In [4], Charlie and Chris presented the assumption that remote compromise of a vehicle is possible, because they believed that injecting CAN message onto the bus through various entry points in a reliable fashion is possible[4]. In the paper, authors performed a remote attack on the vehicle Jeep Cherok and argued that other similar vehicles are facing the same problem. The authors conduct a list of potential entry points of an adversary, which included passive anti-theft system, tire pressure monitoring system, remote keyless entry/start, Bluetooth, radio data system, Wi-Fi, and Telematics. Authors noted that once an adversary can send CAN message using the vulnerabilities of CAN, physical systems could be controlled via remote exploitation. In the experiment authors presented, vehicle engine can be killed, braking and steering control can be taken by adversary as well. As a result of the work, 1.4 million vehicle recall by Fiat Chrysler Automobiles.

In [21], Woo et al. argued that a running connected car can be the target of an remote attack, and to prove their idea, authors demonstrated an actual remote attack on a connected car. The experiment which the authors conduct consisted of two phases. In the first phase, adversary obtains the CAN data frame information through the diagnosis tool connecting to the vehicle. Then the adversary distributed the malicious application masquerades as a normal self-diagnosis app. Note that using a normal self-diagnosis app and a OBD2 scan tool which is connected to the driver's phone using Bluetooth, a driver can obtain the vehicle status. In the second phase, once a driver downloaded the malicious application, then the application server and the driver's phone is connected, the adversary can inject malicious CAN data frame into the vehicle bus.

In [38], P et al. implemented replay attack on CAN environment they constructed, which contains three nodes. In the experiment, one node act as a malicious node and perform the replay attack. Two types of replay attack are demonstrated, one is re-transmitting the whole message, the other one is re-transmitting the message expected the identifier field. Authors noted that the partial replay attack can be difficult to detect, since the identifier field is changed by the adversary, and there is no mechanism on CAN to verify it.

In [40], Murvay and Groza discovered that the vulnerabilities of SAE J1939 standard specification, and noted that various attack(e.g. DoS, impersonation) will be possible on vehicles CAN bus with the standard J1939. In order to prove their point, authors implemented a simulation experiment of sub-network of six nodes based on CANoe with J1939 system. In their experiment, following attacks have been conduct, address claim DoS, request for address claim DDoS, transport protocol DoS, and connection hijacking.

TABLE I
SUMMARY OF SECURITY APPROACHES ON IN-VEHICLE NETWOKR

| Work | Features | Implementation Environment | Limitation |
|------|----------|---------------------------|------------|
| [21] | 1. Demonstrated a remote attack. 2. Authentication and key management scheme. | CANoe[43], DSP-F28335 | 1. Not mentioning pre-shared key derivation. 2. Weak against man-in-middle attack. 3. Partial depending on centralized security node. |
| [16] | 1. Authentication scheme based on a centralized node(Tesla-like). 2. Strong against masquerade and DoS attack. | CANoe, Raspberry Pi 3, Aruduino Zero | 1. Not mentioning pre-shared key derivation. 2. Depending on centralized security node. |
| [10] | 1. Diffie-Hellman key exchange with tree structure and signature. 2. Strong against man-in-middle attack. 3. Testing various elliptic curve for asymmetric cryptography. | Infineon TC224, TC277, TC297, SAM V71 XULTRA | 1. Depending on centralized security node. 2. No efficient session key refreshing function. 3. No formal security analysis. 4. High demand on computing resources. |
| [12] | 1. Key management scheme based on identity-based encryption and broadcast encryption. 2. Optimized operations by parallel computation on FPGA platform. 3. Provide formal security analysis. | IVNS Simulator[29] STM32F415, iMX6, Artix-7 XC7A200T | 1. Weak against man-in-middle attack. 2. Partially depending on centralized security node. 3. High demand for computing resources. |
| Ours | 1. Efficient key management scheme using asymmetric cryptography in only one phase. 2. Provide formal security analysis and security property analysis based on NIST 800-57[27]. | IVNS Simulator, Arduino UNO R3, WinnerMicro W806, STM32F407 | 1. Weak against man-in-middle attack. 2. Partially depending on centralized security node. |

B. *Research on Security Mechanisms for In-Vehicle Networks*

Various researches have been conduct on adding security mechanism on in-vehicle networks[6]-[24], [27], [30], and [34]. Currently, we can categorize solution on in-vehicle network into two groups, one is the cryptography approach, and the other one is intrusion detection system(IDS). The function of IDS is to identify the abnormality occurring on in-vehicle network. Researches on IDS for vehicle security(e.g. [6]-[9], and [27])are using machine learning or statistic approach to detect the suspicious traffic. The main limitations of IDS approach are as follows. For approaches with machine learning, the computation resources requirement may be too high for the ECU connected to the CAN bus. Other than that, installing IDS in each ECU is usually required, this can cause incompatibility and deployment cost problem[39]. Our main focus will be on the cryptography approaches, despite that they are not flawless as the IDS approach, next we will present some of the work on building security protocol on in-vehicle with cryptography approach.

In [21], Woo et al. as mentioned before, designed and illustrated an actual remote attack on a connected vehicle, then in their work, they proposed a novel security protocol for CAN. The feasibility and performance are evaluated according to the experiment that authors conduct. The protocol contains phases of loading pre-shared secret, initial session key distribution, encryption and authentication of the data frame, updating session key, and sharing the session key with external device. In the experiment part, authors conduct the hardware evaluation and software plus hardware evaluation. In hardware evaluation, a DSP-F28335 micro controller is used to act as the secure ECU, which perform the cryptography algorithms. The execution time of them are measured. In software plus hardware evaluation, CANoe is used to simulate the CAN bus. The communication response time, initial session key distribution time, and key update time are tested. In [20], Woo et al. extended their previous work [21] by designing the security protocol considering the ISO 26262 Automotive Safety Integrity Level. ECUs are divided into four groups, each group has different level of security requirement. The lowest only needs CRC, the second level needs authentication, the third level needs authentication plus confidentiality, and the fourth level needs all of above plus external access control.

In [16], Jo et al. proposed a Tesla-like authentication protocol, which contains four phases and a centralized ECU as the authenticator is required. In the first phase, each ECU will derived session key from communication with authenticator, note that the the session key is unique for each pair of ECU and authenticator. Then in the second phase, ECU will broadcast message with MAC derived from the session key. In the third phase, the authenticator will verify the received message, if it fails the authentication, then a report message will be broadcast by the authenticator, if it passes the authentication, then no message will be broadcast. ECU will drop the message if it receives the report message later, otherwise it will not. In the final phase, ECU and the authenticator will update their session key and parameters. In [41], Groza et al. also proposed a Tesla-like protocol, which requires a centralized ECU as the authenticator as well. In this protocol, once a ECU broadcast a message with MAC, the authenticator will verify the message, but unlike [16], if it is correct then a message contains the authentication information will be broadcast, if not no message will be broadcast. However, original Tesla protocol does not mention using a centralized node[42].

Among most of the work on in-vehicle network security, we had the observation that symmetric encryption and MAC calculation are necessary, and thus shared key is necessary. However, in recent years, only few works have been focus on derivation of shared key on CAN without the pre-shared secret. In [10], Musuroi et al. proposed a protocol for group shared key derivation. Authors build the foundation with base Diffie-Hellman key exchange scheme of Station to Station protocol. A centralized trust node will be responsible for signing each message to prevent man-in-middle attack. Then author apply the tree structure into the process of key exchanging. In [12], Carvajal-Roca et al. proposed a key management framework based on identity-based encryption and identity-based broadcast encryption. In their protocol, there has four phases in total. First phase is performed by the vehicle manufacture company, in this part the method for setting up the security parameters on ECU is provided. In the second phase and third phase, a centralized node served as the security module. In these two parts, ECU will perform mutual authentication with the security module and then derive the shared secret key. In the last phase, ECU will derive the session key using identity-based broadcast encryption. Note that, in [12], Carvajal-Roca et al. employ public key encryption in the three main phases for session key derivation, including authentication, initialization and session key phase; in [10], Musuroi et al. also employ asymmetric cryptography in all phases for session key derivation, including share compute, extract & confirm, and verify confirmation phase. On the contrary, we employ the KEM in only one phase and instead use symmetric cryptography plus hash function in subsequent phases for more simplicity and efficiency. To the best of our knowledge, we argue that we are the first to utilize the asymmetric cryptography in only one phase in a key management scheme without long-term pre-shared key.

## III. BACKGROUND AND SYSTEM MODEL

In this section, we first introduce the basic concept of secure key encapsulation mechanism. Then, we present the system model and adversary model for our scheme.

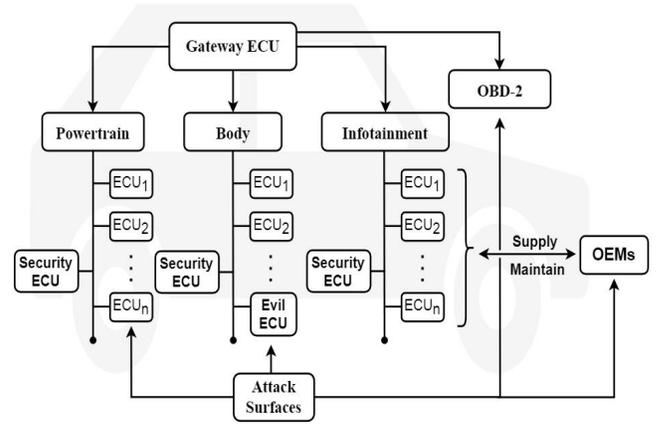

Fig. 1. In-vehicle network with CAN bus protocol. ECU are divided into various groups connecting with security ECU.

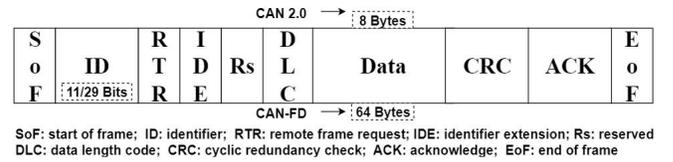

Fig. 2. Standard CAN 2.0 & CAN-FD frame.

### A. Secure Key Encapsulation Mechanism

In the cryptosystem with the public-key setting, a key encapsulation mechanism(KEM) is used to allow sender to run a encapsulation algorithm to derive a random session key and a corresponding cipher text(receiver's public key is know by sender). Receiver can derive the same session key using the decapsulation algorithm on the cipher text. Finally, both end of the user can share a random session key used for symmetric encryption later on[44].

KEM is usually implemented with Diffie-Hellman key exchange mechanism, including our scheme. In this case, receiver's public key consists of two parts, group element $g^a$ and secret key of the random index a($g$ is the generator). Sender perform the key encapsulation by computing the cipher text as $g^b$ for random b. Then, with the key agreement the derived shared session key is $g^{ab} = (g^a)^b$. Different KEMs are implemented under different assumption, in [12], their protocol is under Bilinear Diffie-Hellman Assumption and General Diffie-Hellman Exponent Assumption, in contrast,our work is implemented under Gaped Hashed Diffie-Hellman Assumption.

### B. System Model

In a modern automotive, the vehicular system is divided into four domains depending on their functionality and real-time data requirement. The power train domain, the chassis domain, the body domain, and telematics and infotainment domain are the four main parts of the system. Certain number of ECUs are deployed on the system to collecting monitor data or control certain function. There protocols are used in the in-vehicle network for ECU communication: CAN, Flexray, and LocalConnect Network(LIN). Among these protocols, CAN bus protocol is the most widely used one, and is used for the critical functions, such as power train, braking, and etc.

However, many vulnerabilities exist in CAN bus protocol, this leads to many attacks on in-vehicle network, as previously mention. According to the CAN bus protocol standard[45], we can summarize following problems:

- Broadcast transmission: Once any message is sent by the sender, all nodes connected on the CAN bus can received the message. This might leads to sniffing attack, causing the privacy leaking or other serious threats.

- No authentication mechanism: As shown in figure 2, in CAN message frame, there is no field design for security mechanism. The message is transmission in plain text and without source and destination address. This might leads to replay attack and masquerade attack.

- Arbitration mechanism: When multiple nodes are sending messages, determined by the data of identifier field of each message, the smallest one wins the highest priority. Because of this, an adversary can stop the normal communication by sending message with highest priority. This might leads to DoS attack.

The components in our scheme are summarized as follows, and figure 1 shows the CAN bus protocol on vehicle.

- CAN Bus: We implemented our security scheme on CAN bus, note that it is also compatible for CAN-FD, which is the second generation of CAN. The bus message transmission bit rate ranges from 125kb/s to 1Mb/s, we find that most of the work consider to be 500kb/s, thus we will follow as well.

- Original Equipment Manufacturers(OEM): OEMs produce and install auto parts during the construction of the vehicle. It is reasonable to design the protocol that, OEMs run the algorithm to store security parameters on ECU. For the functionality of OEMs, our system model is built similar to that in [12], [22].

- Security ECU(SECU): The security ECU is a central node, responsible for authentication and key derivation. The main benefit of deploying a high-end central node is reduction on the computation time of performing complex cryptography algorithm. In fact, most work on in-vehicle network security are using central security ECU for authentication.

- ECU: Grouping certain number of ECU into a group is natural, since vehicle system is divided into various domain, so as ECU is group into the specific domain in a sense of way. In a modern vehicle, the number of ECU can be up to 80 and even 150 in some luxury cars[46].

To eliminate the situation that the adversary gains the control of certain ECU and thus finds the security parameters. We will argue that, as suggested in [14] and [47], in ECU and SECU, all these critical data can be stored in tamper-resistant memory such as the Trusted Platform Module (TPM)[48].

*C. Adversary Model*

Our proposed scheme is built on the assumption that the adversary is able to connect to the vehicle through various attack surfaces previously mentioned and perform certain types of attack by taking advantage of the vulnerabilities of the in-vehicle network, as shown in figure 1. The three attack surfaces will be described as below:

TABLE II
PROPOSED SCHEME ARCHITECUTRE

| Phase | Description | Central Node |
|---|---|---|
| Initialization | Generate and load security parameter | Not Needed |
| Pairwise Secret Generation | Generate pairwise secret between ECU and SECU | Needed |
| Group Shared Secret Generation | Generate group shared secret among ECU | Needed |
| Session Key Generation | Generate session key among ECU | Not Needed |
| Session Key Refreshing | Update session key for new round | Not Needed |

(a) Malicious ECU & OBD2: An adversary is able to place malicious ECU in the CAN and connect to the CAN remotely using malicious app as described in above. In both cases, the adversary can then perform various type of attacks such as sniffing attack, replay attack, and etc.

(b) OEM: An adversary is able to gain the information that the OEM have. In this case, the adversary can perform attacks with the knowledge of the security parameters.

In this scenario, the adversary has the ability to achieve following: inject malicious messages, complete or partial re-transmission of the messages, eavesdropping on messages transmission on in-vehicle network. Note that, as other protocol's assumption[12] - [13] and [20] - [21], our scheme is not design for firmware modification, and other than that, man in the middle attack is not prevented either although we have the compensation for it in certain degree, this will be explained later on section IV.

IV. THE PROPOSED KEY MANAGEMENT SCHEME

In this section, we presented our proposed novel key management scheme on in-vehicle network and explain the scheme in detail. Then we presented the correctness analysis and security analysis of our scheme according to the standard of NIST 800-56A and NIST 800-57. Table III presents the notations used on our propose scheme.

Our proposed scheme consists of five phases, as shown in table II. To have a preliminary understanding, here we roughly discuss the scheme. In the first and second phases, inspired by the work of Kiltz [31], an elliptic curve cryptography based key encapsulation mechanism is utilized for the pair-wise secret derivation between each ECU and SECU. Then in the third phase, SECU and each ECU can make use of the pair-wise secret to further derive the group shared secret, where encryption and authentication and key derivation algorithm are capitalized. Next in the fourth phase, now the scheme does not rely on the SECU anymore, an arbitrary ECU can be responsible for the role of sender. This chosen ECU can derive the session using the random seed and then broadcast the seed after authentication so that other ECU can do the same. In the last phase, each ECU can update the session key on their own.

TABLE III
NOTATION ON PROPOSED SCHEME

| Notation | Description |
|---|---|
| $ECU_i$ | Electronic control unit with the identity of **i** |
| SECU | Centralized Secure ECU |
| CTR | Message counter |
| $K_i$ | Shared secret between ECU ID of **i** and SECU |
| SK | Group shared secret among ECU and SECU |
| $SSK_R$ | Session key of round **R** |
| H | Hash function |
| HMAC | Derivation function of keyed-hash message authentication code |
| HKDF | Key derivation function based on hash function |
| $Encry_{K_i}$ | Symmetric encryption function using key of **i** |
| $Decry_{K_i}$ | Symmetric decryption function using key of **i** |

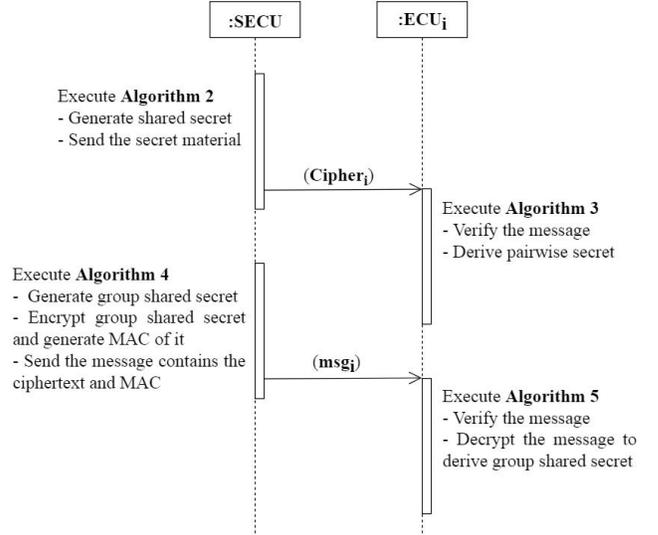

Fig. 3. Sequence diagram for phase 2 & 3.

*A. Proposed Scheme*

*1) Initialization:* In this phase, we present the algorithm to generate security parameters when the ECU and SECU are being constructed. As suggested in [12] and [22], we argue that it is reasonable to assign this process to OEMs, as illustrated in Figure 1.

OEMs will execution ParameterGeneration(), as described in Algorithm 1, to derive the security parameters. When the algorithm is finished, security parameters will be assigned to every ECU respectively, and SECU will store the public key of other ECU. As suggested in [14] and [47], we also suggest that, these critical security information can be stored in the tamper-resistant memory of a security module such as the TPM[47]. Typically, a TPM can provide following functions: proving an ECU identity, reporting the software in use, and remote deployment of maintenance updates, cryptosystem key establishment, storage, management and use[50]. Note that, although the public key and private key are disclosed, the group shared secret still remains secret to OEMs, this provide more obstacles for adversary to hack the vehicle.

---
**Algorithm 1** ParameterGeneration()
---
1: Select a random $k \in N$, then generate prime number **p**, where $2k < p < 2k+1$
2: Select **G**, where **G** is a cyclic group of order **p**, then select a random generator **g** of **G**
3: Choose a random hash function $H: G \to \{0, 1\}^{l(k)}$ that outputs $l(k)$ bits, where $l()$ is a fixed polynomial
4: **for**(i == 0; i < the number of ECU; i++)
5:    Select random $x_i, y_i \in Z_p^*$
6:    Compute $g^{x_i} \to u_i$; $g^{y_i} \to v_i$
7:    Set $(u_i, v_i) \to pk_i$; $(x_i, y_i) \to sk_i$
8:    Output $(G, g, p, H, sk_i, pk_i)$
9: **end for**

*2) Pairwise Secret Generation:* In this phase, we present the algorithms for generating the shared secret between each ECU and SECU. During this phase, SECU will execute PairwiseSecretGeneration(), as described in Algorithm 2, to derive the pairwise shared secret $K_i$ with each ECU and send the shared secret material **r** after encapsulation, then each ECU can decapsulate the message and then derive the pairwise shared secret $K_i$ by executing PairwiseSecretDecapsulation(), as described in Algorithm 3.

---
**Algorithm 2** PairwiseSecretGeneration()
---
1: Load parameters and extract $pk_i \to (u_i, v_i)$
2: **for**(i == 0; i < the number of ECU; i++)
3:    Select random $r \in Z_p^*$
4:    Compute the following:
    $g^r \to c_i$; $H(c_i) \to tem$; $(u_i^{tem} v_i)^r \to \alpha_i$
5:    Derive the shared secret:
    $H(u_i^r) \to K_i$
6:    Set $(c_i, \alpha_i) \to Cipher_i$
7:    Send $Cipher_i$ to $ECU_i$
8: **end for**

---
**Algorithm 3** PairwiseDecapsulation($Cipher_i$)
---
1: Receive $Cipher_i$ from SECU and extract $(c_i, \alpha_i)$
2: Load parameters and extract $sk_i \to (x_i, y_i)$
3: Compute the following:
    $H(c_i) \to tem'$; $c_i^{x_i \cdot tem' + y_i} \to \beta_i$
4: **if** $\alpha_i == \beta_i$, **then**
5:    Compute $H(c_i^{x_i}) \to K_i$
6: **else** terminate the process
7: **end if**

*3) Group Shared Secret Generation:* In this phase, we present the algorithms to derive group shared secret **SK**, as described in Algorithm 4 and Algorithm 5. During this phase, SECU executes algorithm 4 GroupSharedSecretGeneration() to generate the group shared secret **SK**. At first SECU randomly generate the group shared secret **SK**, then use HKDF to derive two keys $(K_{i\_1}, K_{i\_2})$ for every ECU in the group, next send the secret to them after encryption and authentication with MAC. Each ECU in the group receive the message and derive the group shared secret **SK** by executing algorithm 5 GroupSharedSecretVerify().

In phase 2 and phase 3, which includes pairwise secret generation phase and group shared secret generation phase, SECU and regular ECU each is assigned with two algorithms and two messages are dispatched in total, as a result, the group shared secret is derived by every ECU, as shown in figure 3.

---

**Algorithm 4** GroupSharedSecretGeneration()
1: Select a random group shared secret **SK**
2: **for**(i == 0; i < the number of ECU; i++)
3:   Compute the following:
4:   $HKDF(K_i) \to (K_{i\_1}, K_{i\_2})$
5:   $Encry_{K_{i\_1}}(SK) \to m_1$; $HMAC(m_1, K_{i\_2}) \to m_2$
6:   Set $msg_i = (m_1, m_2)$, then send $msg_i$ to $ECU_i$
7: **end for**

---

**Algorithm 5** GroupSharedSecretVerify($msg_i$)
1: Receive $msg_i$ from SECU and extract $(m_1, m_2)$
2: Compute $HKDF(K_i) \to (K_{i\_1}, K_{i\_2})$
3: Compute $HMAC(m_1, K_{i\_2}) \to m_2^*$
4: **if** $m_2^* == m_2$, **then**
5:   Compute $Decry_{K_{i\_1}}(m_1) \to SK$
6: **else** terminate the process
7: **end if**

---

*4) Session Key Generation:* In this phase, we present the algorithm to generate group shared session key in the first round, as described in algorithm 6 and algorithm 7. Note that in the last two phases, standard ECU will be sufficient. During this phase, any ECU can be responsible for executing SessionKeyGeneration() in order to generate the first session key and broadcast it to all the ECU in its group. Then rest of the ECU in the group execute SessionKeyVerify() in order to verify the authenticity of the broadcast message then generate the first session key for message authentication.

---

**Algorithm 6** SessionKeyGeneration()
1: Select a random secret **seed**
2: Set the round number counter **R** as zero
3: Compute the following:
  $HKDF(SK) \to (SK_1, SK_2)$
  $HKDF(seed, R, SK_1) \to SSK_R$
  $HMAC(seed, SK_2) \to mac$
4: Set $(seed, mac) \to msg$, then broadcast **msg** to others

---

**Algorithm 7** SessionKeyVerify(**msg**)
1: Receive **msg** extract $(seed, mac)$
2: Set the round number counter **R** as zero
3: Compute the following:
$HKDF(SK) \to (SK_1, SK_2)$; $HMAC(seed, SK_2) \to mac'$
4: **if** $mac' == mac$, **then**
5:   Compute $HKDF(seed, R, SK_1) \to SSK_R$
6: **end if**

---

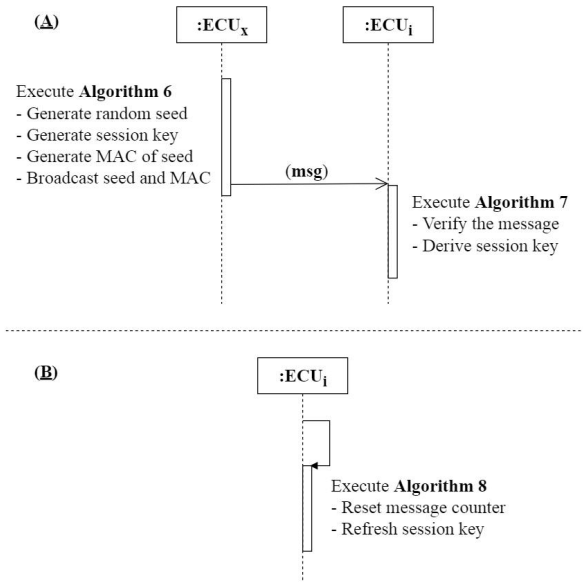

Fig. 4. Sequence diagram for phase 4 & 5.

*5) Session Key Refreshing:* In this phase, we present the algorithm to update session key when the message counter reaches its limit, as described in algorithm 8. Note that in this phase, communication among ECU is not needed.

---

**Algorithm 8** SessionKeyRefresh(**CTR**)
1: **if CTR** == maximum, **then**
2:   Set **CTR** as zero
3:   Set $R \to R + 1$
4:   Compute $HKDF(SSK_{R-1}, R) \to SSK_R$
5: **end if**

---

In phase 4 and phase 5, which includes session key generation phase and session key refreshing phase, any regular ECU(e.g. ECU with ID of x) is capable of the following task, generating random seed and derive session key using the seed, then generate MAC of seed, finally broadcast the message which contains the seed and MAC to the group it belongs. All ECU(except ECU with ID of x) in the group receive the message and verify it, then derive the session key, as shown in part A of figure 4. Regular ECU increases the message counter CTR by one every time it broadcast a data frame, when CTR reaches its limit, ECU derives new session key without any message being sent, as shown in part B of figure 4.

*B. Protocol Correctness Analysis*

In this session, we present the correctness analysis for the cipher message Cipher and pairwise key K in phase 2. Recall that the key pairs are constructed for asymmetric cryptography, namely $(u, v) \to pk$ and $(x, y) \to sk$. It is reasonable to argue that $(c, \alpha) \to Cipher$ is consistent when the following condition is satisfied, when $H(c) = tem$:

$$c^{x \cdot tem + y} = \alpha \qquad (1)$$

When the algorithm is being executed correctly, the cipher message will be generated as follows.

$$(g^r, u^{r \cdot tem} v^r) = (c, \alpha) = \text{Cipher} \quad (2)$$

And thus we can naturally see that.

$$\alpha = (u^{tem} v)^r = (g^{x \cdot tem + y})^r = c^{x \cdot tem + y} \quad (3)$$

In this way we consider Cipher is consistent. Then in the case of pairwise key derivation, the sender and the receiver, namely the central secure ECU and the regular ECU in the group, can perform encapsulation and decapusaltion in order to derive the pairwise secret as follows.

$$H(c^x) = H((g^r)^x) = H(u^r) = K \quad (4)$$

Here, the correctness of the asymmetric cryptography in our protocol is shown.

*C. Protocol Security Analysis*

In this section, we present the security analysis of our protocol. Recall that in the second phase, a KEM is used to generate pairwise key, which means that chosen-ciphetext attack will be the main threat to our scheme. Thus we divide the analysis into three parts. In the first part, a formal security proof for the KEM in our scheme is presented. Then in the second part, we discuss the security property of our protocol in general and in terms of the NIST guidelines [25] and [26]. At last we discuss the remaining possible attack scenarios.
- Chosen-Ciphertext Attack

In this subsection, we present the security proof for the KEM, which is using asymmetric cryptography and based on Gap Hashed Diffie-Hellman(GHDH) assumption, utilized in our proposed scheme. A more comprehensive proof can be found on [31].

First, we summarize the security property of a regular KEM. In a KEM scheme, there will be three polynomial-time algorithms, denoted as $KEM = (\mathbf{KG}, \mathbf{Encap}, \mathbf{Decap})$, also the key space is define as $KS(k)$. The functions of each algorithm are: $\mathbf{KG}(1^k) \rightarrow (sk, pk)$ a key pair of secret key sk and public pk are generated with respect to the security parameter $k \in \mathbb{N}$. $\mathbf{Encap}(pk) \rightarrow (K, C)$ a ciphertext C and a key K are generated with respect to pk. $\mathbf{Decap}(sk, C) \rightarrow (K)$ the ciphertext C is decrypted and the key K is reconstructed using sk. Assume there is an adversary A aims to break the KEM with the privilege of accessing $\mathbf{Encap}()$ and $\mathbf{Decap}()$, we conduct an experiment as follows. Note that adversary A's goal is to distinguish between a random string and a correctly generated key.

$$\mathbf{Exp}_{KEM, A}^{KEM-CCA}(k)$$
$$\mathbf{KG}(1^k) \rightarrow (sk, pk)$$
$$KS(k) \rightarrow (K_0); \mathbf{Encap}(pk) \rightarrow (K_1, C)$$
$$\{0, 1\} \rightarrow \alpha$$
$$A^{Decap()}(pk, K_\alpha, C) \rightarrow \alpha^*$$
If $\alpha^* = \alpha$ then return 1, else return 0

Note that $K_0$ is randomly generated from the key space and has the same length as $K_1$. Adversary A is able to query the decapsulation algorithm using ciphertext $C^*$ as input, however with the only restriction that using C is not allowed. Then the advantage of adversary A in the experiment can be defined as the distance from $\frac{1}{2}$ to the probability of adversary A determined whether $K_\alpha$ is a random string or correctly generated key, as shown below.

$$\mathbf{Adv}_{KEM, A}^{KEM-CCA}(k) = \left| \Pr[\mathbf{Exp}_{KEM, A}^{KEM-CCA}(k) = 1] - \frac{1}{2} \right| \quad (5)$$

A KEM is considered to be indistinguishable against chosen-ciphertext attacks(IND-CCA) in the case of $\mathbf{Adv}_{KEM, A}^{KEM-CCA}(k)$ is a negligible function[51] in k for polynomial time adversary A.

Second, we summarize the security property of the target collision resistant hash functions(TCR). In a normal condition, a TCR can be shown as: $G \rightarrow \mathbb{Z}_p$, where G is a cyclic group of order p. Assume there is an adversary H aims to break the TCR, namely finding the hash collision, the advantage of H can be defined as below.

$$\mathbf{Adv}_{HASH, H}^{HASH-TCR}() = \Pr[(\mathbf{TCR}(t) = \mathbf{TCR}(t^*)) \cap (t \neq t^*)] \quad (6)$$

Third, we summarize the security property of the GHDH assumption. In order to explain this, a scheme under GHDH assumption will be shown. The scheme is first parameterized by a parameter generation function $\mathbf{GEN}(k) \rightarrow (G, g, p, H)$. Using the input of k, the function output a cyclic group G of prime order p, where $2^k < p < 2^{k+1}$, a random generate g from G, and at last a random hash function H: $G \rightarrow \{0, 1\}^{l(k)}$, which output string with length of $l(k)$ bits. Assume there is an adversary B aims to break the scheme, namely distinguish between the result $H(g^{x \cdot y})$ and a random string with the same length, in which x and y are randomly selected from $\mathbb{Z}_p$, we conduct an experiment as follows.

$$\mathbf{Exp}_{GEN, H, B}^{GHDH}(k)$$
$$\mathbf{GEN}(k) \rightarrow (G, g, p, H) = HG$$
$$\mathbb{Z}_p \rightarrow (x, y); \{0, 1\}^{l(k)} \rightarrow W_0; H(g^{x \cdot y}) \rightarrow W_1$$
$$\{0, 1\} \rightarrow \alpha$$
$$B^{DDHsolve()}(k, HG, g^x, g^y, W_\alpha) \rightarrow \alpha^*$$
If $\alpha^* = \alpha$ then return 1, else return 0

Note that the adversary B are able to use Decisional Diffie-Hellman(DDH) oracle DDHsolve(). Then the advantage of adversary B can be defined as the distance from $\frac{1}{2}$ to the probability of adversary B determined whether $W_\alpha$ is a random string or correctly generated hash value, as shown below.

$$\mathbf{Adv}_{GEN, H, B}^{GHDH}(k) = \left| \Pr[\mathbf{Exp}_{GEN, H, B}^{GHDH}(k) = 1] - \frac{1}{2} \right| \quad (7)$$

The sample scheme based on the GHDH assumption is considered to be tenable when $\mathbf{Adv}_{GEN, H, B}^{GHDH}(k)$ is a negligible function[51] in k for polynomial time adversary B.

Next, we present the security analysis of the KEM used in our proposed scheme, namely that the following discussion is in the context of the KEM constructed as Algorithm 1, 2 and 3. Assume there is an adversary A aims to break the chosen-ciphertext security of the KEM scheme with the non-negligible advantage of $\mathbf{Adv}_{KEM, A}^{KEM-CCA}(k)$. Besides, there is an adversary B aims to solve the GHDH problem with the advantage of $\mathbf{Adv}_{GEN, H, B}^{GHDH}(k)$. Other than that, there is an adversary H aims to break the hash function with the advantage of $\mathbf{Adv}_{HASH, H}^{HASH-TCR}()$. In this case, adversary B is capable of running adversary A and H simultaneously, namely the adversary is able to exploit the SECU and ECU. In the end, the goal of adversary A is to distinguish between the correctly generated key K and a random string, the goal of adversary B is to distinguish between the correctly generated hash result W and a random string, and the goal of adversary H is to find the hash collision. Then we conduct an experiment as follows.

**Key Generation**: Firstly, adversary B set up the security parameters via executing function **GEN**$(k) \to (HG, g^x, g^y)$, and let $u = g^x$, $W_0$ be a random string and $W_1 = H(u^y)$, where $|W_0| = |W_1|$. Note that, adversary A and H can be executed by adversary B simultaneously.

**Challenge**: Then, a random value $r^* \in Z_p$ is selected by adversary B. The challenge ciphertext can now be defined as $(g^y, (g^y)^{r^*}) \to (c^*, \alpha^*) \to Cipher^*$, generated key $K^* = W_i$, where $i \in \{0, 1\}$, associated tag on the challenge ciphertext $H(c^*) \to tem^*$. In all, a query of $(pk, Cipher^*, K^*)$ is sent to adversary A by adversary B.

**Decapsulation**: Recall that in algorithm 2, $pk = (u, v)$, according to [31], we can calculate the v as: $u^{-tem^*} g^{r^*} \to v$. In this way, originally $\alpha = (u^{tem} v)^r$ can be calculate as:

$$(u^{tem} v)^r \to (u^{tem} u^{-tem^*} g^{r^*})^r \to (u^r)^{tem - tem^*} c^{r^*} \to \alpha \quad (8)$$

Then, the original $K = H(c^x)$ can be alternately calculated as:

$$K = H(c^x) \to H(u^r) \to H\left(\left(\frac{\alpha}{c^{r^*}}\right)^{(tem - tem^*)^{-1}}\right) \quad (9)$$

Note that, adversary A is able to execute the decapsulation oracle **DecapO()**, using the input of a random generated string of $Cipher \to (c, \alpha)$, with the restriction that Cipher is not the same as $Cipher^*$. Before the K is calculated, adversary A should perform a check via calculating $H(c) \to tem^*$. In this way, a total of three cases will occur as follow:

(a) When $(c = c^*) \cap (tem = tem^*) = 1$, the query is illegal and therefore discarded by adversary A, because this is against the rule mentioned above.
(b) When $(c \neq c^*) \cap (tem = tem^*) = 1$, adversary A found the collision in hash function, the query is also illegal and therefore discarded.
(c) When $(tem \neq tem^*) = 1$, the query is accepted, then the key K is calculated as described in (9).

By now, the behaviour of the decapsulation oracle executed by adversary A is stated.

**Guess**: Before the end of the program, adversary A will disclose its guess of the key $K^*$ as $\sigma \in \{0, 1\}$, in which when $\sigma = 1$ indicates that $K^*$ is the correctly generated key, and otherwise $K^*$ is a random string. Similarly, adversary B will also disclose its guess of the W as $\rho \in \{0, 1\}$, in which when $\rho = 1$ indicates that W is the correctly generated hash result, and otherwise W is the random string.

By now, the behaviour of adversary B is stated.

**Analysis**: Now we define **WIN** as the event that adversary B successfully break the KEM. The advantage of adversary B can be defined as:

$$\mathbf{Adv}_{KEM, B}^{GHDH}(k) = \left|\Pr[\mathbf{WIN}] - \frac{1}{2}\right| \quad (10)$$

It is obvious to see that the advantage of adversary B is associated with the advantage of adversary A and H. Note that the advantage of adversary H, namely H finding the collision in hash function, is stated as (6), and the advantage of adversary A, namely A making the correct guess, is stated as (5). As mentioned in [31], we can now alternately define the advantage of adversary B as:

$$\left[\mathbf{Adv}_{KEM, A}^{KEM-CCA}(k) - \mathbf{Adv}_{HASH, H}^{HASH-TCR}()\right] \leq \mathbf{Adv}_{KEM, B}^{GHDH}(k) \quad (11)$$

TABLE IV
CRYPTOPERIODS IN OUR PROPOSED SCHEME

| Key Type | Used In Phase | Cryptoperiod |
|---|---|---|
| Private key agreement key | Phase 1 & 2 | Up to 2 Years |
| Public key agreement key | Phase 1 & 2 | Up to 2 Years |
| Symmetric authentication key | Phase 3, 4 & 5 | Up to 2 Years |
| Symmetric data encryption key | Phase 3 | Up to 5 Years |

TABLE V
SECURITY STRENGTH IN OUR PROPOSED SCHEME

| Cryptography Algorithm | Used In Phase | Security Strength |
|---|---|---|
| Symmetric Key Algorithm (AES-128) | Phase 3 | 128 Bits |
| Elliptic Curve Diffie-Hellman (NIST K-283, Curve25519, FourQ) | Phase 2 | 128 Bits |
| Keyed Hash Message Authentication Code (HMAC-SHA256) | Except Phase 1 | 256 Bits |
| Keyed Hash Key Derivation Algorithm (HKDF-SHA256) | Phase 4 & 5 | 256 Bits |

Hence, under the assumption of GHDH, as state in (11), we argue that the KEM used in our proposed scheme is secure against chosen-ciphertext attacks.

- Security Property Analysis

In this subsection, we present the security property analysis for our proposed scheme in terms of two aspects, namely cryptoperiods and security strength, according to the guideline from [25] and [26].

A cryptoperiods can be roughly defined as the time span that a certain type of key is employed in a cryptosystem. A well defined cryptoperiods can be beneficial to a cryptosystem in various ways: minimizing the time available for an adversary to break the system; minimizing the harm to the system when any key is exposed and etc. In [25], a recommended cryptoperiods sorted by key type is given. Recall that, in our proposed scheme, we employ key types of: private key agreement key, public key agreement key, symmetric authentication key, and symmetric data encryption key. Then combine the guideline of [25] with our proposed scheme, we state the cryptoperiods of each part in TABLE IV. We argue that although the cryptoperiods are up to two years or more, based on the consideration of safety and convenient, changing the whole set of keys used in the protocol once a year together with the annual vehicle inspection will be suggested.

The security strength(security level in some articles) of a certain type of cryptosystem can be defined as the amount of operation that an adversary need to break the system. More specific, usually, the security strength is defined in bit from the set $\{112, 128, 192, 256\}$. For the propose of safety, we employ the approved cryptography algorithms from guideline of [25] in our proposed protocol. However, with the greater security strength, comes with the greater time consumption, finding the algorithm with the best balance will be suggested. In our proposed protocol following algorithms are employed, the symmetric key algorithm of AES-128, the elliptic curve based Diffie-Helleman of curve FourQ[52] and others, HMAC and HKDF based on various hash functions. The security strength in our propose protocol are presented in TABLE V.

- Other Possible Attacks Analysis

In this last subsection, we present the analysis of all the possible attacks on our proposed scheme. Recall that in Section III, we stated that there will be three attack interfaces, namely malicious ECU, OBD2 and the OEM. The analysis blow is given with respect to them.

(a) Malicious ECU & OBD2: If the adversary manage to place in a malicious ECU or getting the access of OBD2, he can then perform various attacks as follows.

For the case of replay attack: Firstly, there will be no benefit for the adversary at all performing this attack. Recall that throughout our proposed scheme, only one-way communication is established, meaning that replay any message cannot acquire any privilege for any further malicious activity. Secondly, this can be solved by reserving the receiving message for a short period of time and when a duplicated message is received, then it can simply discard it.

For the case of sniffing attack: Because there is no any plaintext or explicit key being transmitted in our proposed scheme, the adversary will not manage to gain any useful information in this attack.

For the case of spoofing attack: In our proposed scheme, the integrity of every message is provided according to the hash functions. Except when the adversary manage to break the cryptosystem and find the key, the spoofing attack will not be a problem.

For the case of man-in-the-middle(MIM) attack: To prevent this attack, a trusted third party acted as a certificate authority is necessary in the cryptosystem. However, this will not be practical in the case of CAN in vehicle. To compensate this, according to [53], we argue that our proposed scheme can provide MIM integrity, that if an adversary try to perform a MIM attack, throughout the whole protocol, it must modify every message that is being transmitted or else the attack will be detected.

(b) OEM: Recall that in the first phase of our proposed scheme, OEM is participated and responsible for providing the device with the security parameter embedded. In this case, an adversary will attempt to gain the information of the security parameter. However, even if the adversary manage to success in that, one still cannot derive the pair-wise secret since the parameter randomly generated by SECU remain unknown.

TABLE VI
HARDWARE CHARACTERISTTIC

| Microcontroller | Maximum Frequency | Flash | SRAM |
|---|---|---|---|
| Arduino UNO R3 | 20 MHz | 32 KB | 2 KB |
| WinnerMicro W806 | 240 MHz | 1 MB | 288 KB |
| STM32F407 | 168 MHz | 1 MB | 192 KB |

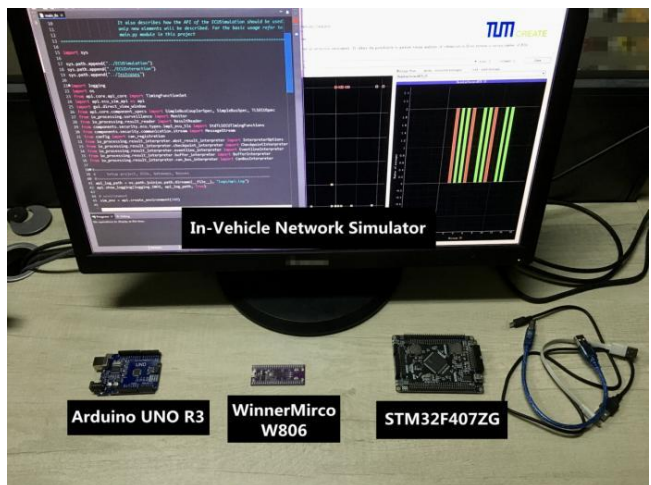

Figure. 5. Experiment Environment Setup

V. IMPLEMENTATION AND PERFORMANCE ANALYSIS

In this section, we present the detail of our experiment and analysis of the performance of our proposed scheme based on the experiment we conducted. Similar to [12], [13], [20] and [21], the experiment is divided into two parts, namely hardware-based experiment and software-based simulation. Figure 5 shows our experiment setup.

*A. Hardware-based Experiment*

In this part, we present the detail of the set up of our hardware-based experiment, then the performance is shown along with the analysis. In this experiment, we used three different microcontrollers(MCUs) as the hardware platforms in order to measure the time consumption for different cryptography algorithms. TABLE VI concludes the characteristics of all the hardware. Among these hardware, STM32 and W806 have a superior computing ability, therefore represented the high-end ECU(SECU and regular ECU); UNO R3 has a very limited computing ability, therefore represented the low-end ECU(ECU with low or no security requirement). More information about the three MCUs used in the experiment can be found at [32], [33] and [54] respectively.

We reiterate here that the purpose of the hardware-based experiment is to measure the time consumption of each cryptography algorithms in our proposed scheme. Now we run back our proposed scheme as follows.

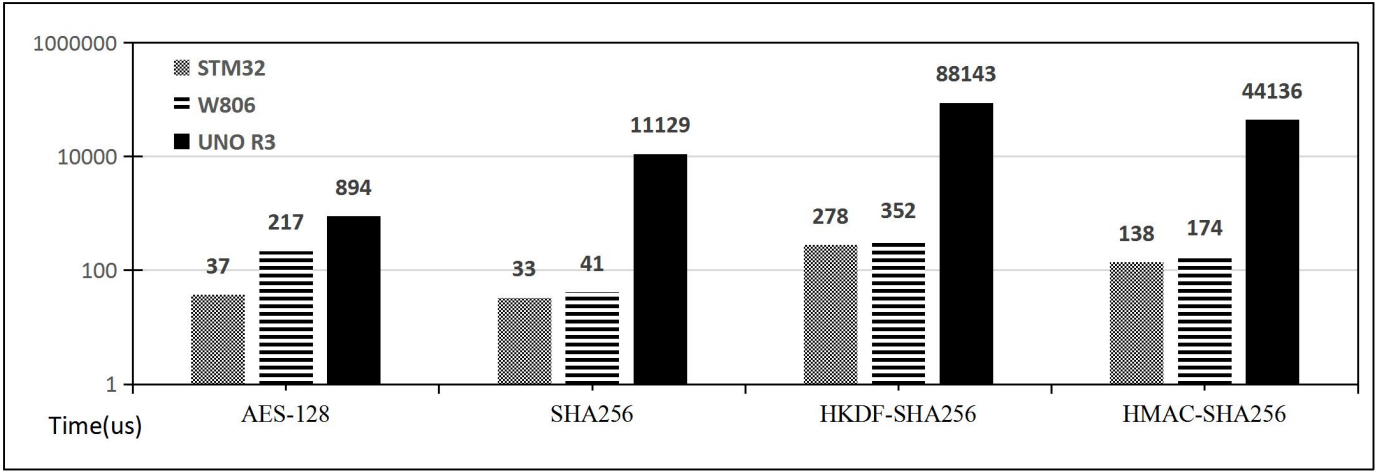

Fig. 6. Time Consumption For Cryptography Algorithms On Three MCUs.

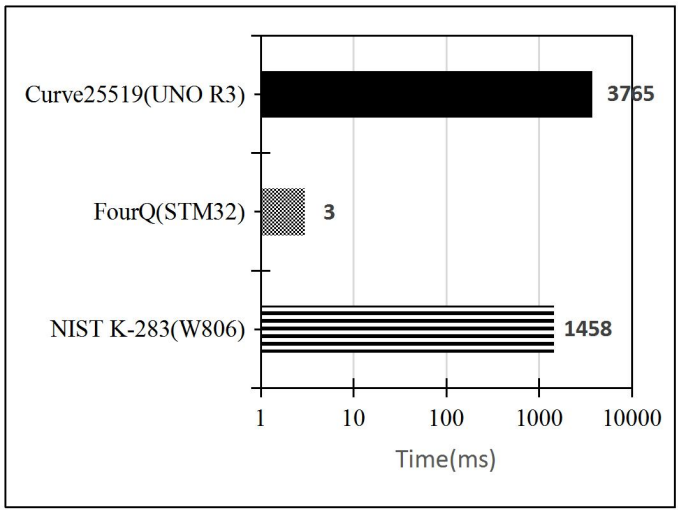

Fig. 7. Time Consumption For ECCDH On Three MCUs.

In phase 1 and 2, a KEM using asymmetric cryptography and hash function, more preciously elliptic curve cryptography based Diffie-Hellman algorithm and SHA256, is constructed in order to derive pair-wise secret, we implemented this using TinyECDH[55](NIST K-283) on W806, Curve25519[56] on UNO R3 and FourQlib[57] on STM32.

In phase 3, 4 and 5, symmetric cryptography and hash functions, more specific AES-128, HMAC-SHA256 and HKDF-SHA256, are employed in order to derive group shared secret and session key, we implemented the experiment using TinyAES[58] and the source code extracted from RFC 6234[49] on all three MCUs.

Note that when we were choosing the specific algorithm to used in our hardware-based experiment, we took these two things into consideration: Firstly, the tradeoff between the security strength and the performance, namely time consumption. Naturally, with the greater security strength, comes with the lower performance. Here, we prefer the one with the best performance yet satisfying sufficient security strength. Secondly, the compatibility of the MCU and algorithm implemented code. It is easier in this case, we simply chose the algorithm has the best performance associated with the specific MCU.

The experiment result of our experiment, namely the time consumption of each cryptography algorithm, can be found at Figure 6 and 7. As shown in Figure 6, all cryptography algorithms except the asymmetric one are tested on the three MCUs. Recall that in Algorithm 4 to 7, AES and HMAC are used to derive the group shared secret **SK**, and HKDF is used to derive the session key **SSK** in the end. Hence we have the result of following: STM32 acquired the best performance using less than 300 µs to accomplish HKDF-SHA256 and no more than 40 µs to complete AES-128 and SHA256; The low-end MCU UNO R3 has the worst performance, it took more than 88 ms to accomplish the hardest task and almost 1 ms to complete the easiest task.

In Figure 7, three ECCDH algorithms were tested on three MCUs respectively. Recall that in Algorithm 2 and 3, in order to derive the pair-wise secret $K \leftarrow H(u^r)$, elliptic curve operations and hash functions were performed. Hence we set up as following: NIST K-283 curve was run on W806, with the result of less than 2 seconds; Then FourQ was run on STM32, with the best result of 3 ms; Lastly, Curve25519 was run on UNO R3, with the worst result of nearly 4 seconds. In the next section, a simulation of vehicle CAN communication using our proposed scheme will be presented combined with the data collected from the hardware-based experiment.

### B. Software-based Simulation

In this part, we presented a software-based simulation in order to measure the performance of our proposed scheme and an analysis of the simulation. Inspired by [12], [13], [20] and [21], we design the simulation based on the performance measured from hardware-based experiment. More specific, before any ECU broadcast a message, time delay is applied representing the time consumption for executing cryptography algorithms in our simulation. The simulation of in-vehicle network(CAN in specific) communication is built based on the open source software IVNS[29]. The IVNS is an open source simulator built on Python, simulates the in-vehicle network communication of CAN and CAN-FD protocol. The main purpose of IVNS is to evaluate the performance of a security scheme and was utilized by various works [12], [17] and [30]. With the help of this tool, we can program the behaviour of the ECU and set up any amount of ECU in a group in the network. Next we present the performance evaluation of our proposed scheme in the simulated CAN environment we constructed. The evaluation is provided based on phase 2, 3 and 4 of our proposed scheme. We exclude the first phase and last phase with following reasons: In phase 1, recall that the initialization of security parameters $(G, g, p, H, sk_i, pk_i)$ was assigned to the OEM; In phase 5, the session key refreshing mechanism is independent of communication, namely each ECU directly update $SSK \leftarrow HKDF(SSK_{R-1}, R)$ without any message being transmitted.

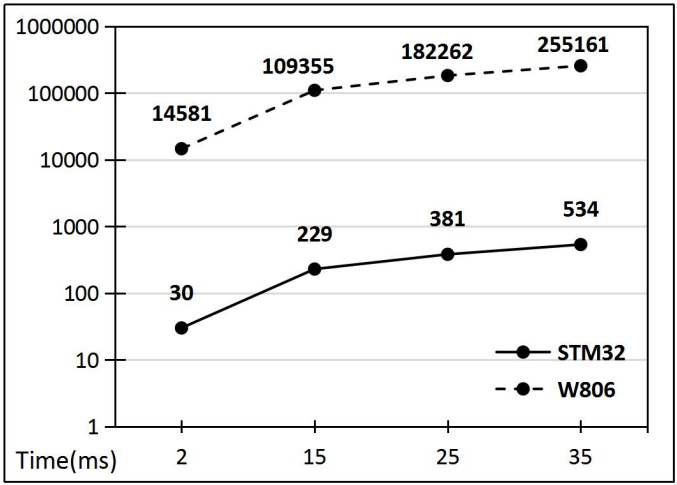

Fig. 8. Execution Time For Phase 2 Based On The Quantity of ECU.

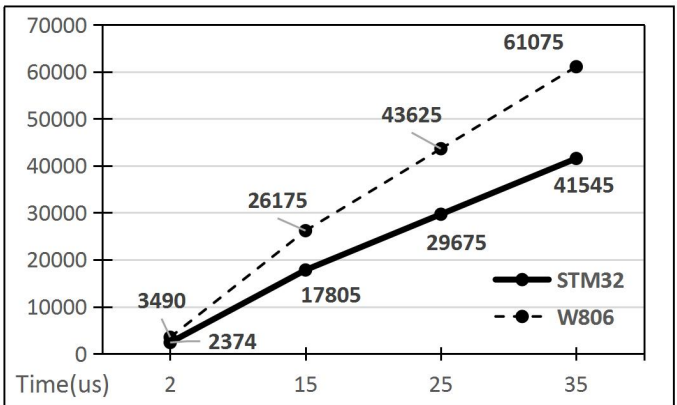

Fig. 9. Execution Time For Phase 3 Based On The Quantity of ECU.

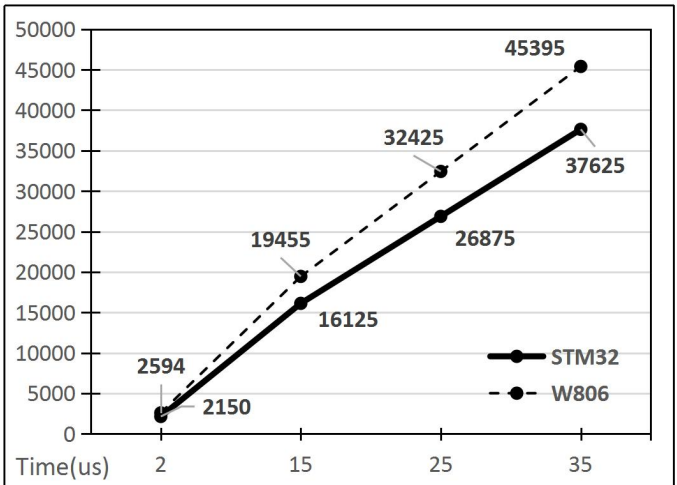

Fig. 10. Execution Time For Phase 4 Based On The Quantity of ECU.

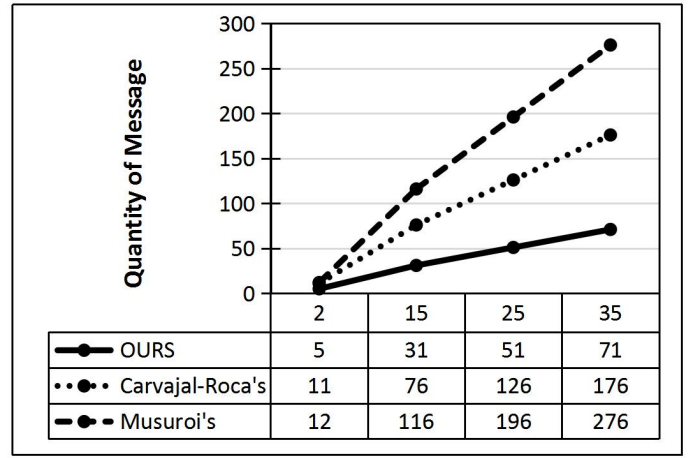

Fig. 11. Total Message Transmission Comparison Based On The Quantity of ECU.

We now present the detail of our simulation environment. The simulated communication among ECUs and SECU is based on CAN-FD protocol in IVNS, with the transmission data rate of 1 Mbit/s and the data field length of 64 bytes in a data frame. We defined various four groups with different size to measure the execution time for phase 2, 3 and 4 in our proposed scheme, they are {2, 15, 25, 35}. Figure 8 shows the execution time for phase 2 pair-wise secret generation. Recall that in phase 2, KEM with ECCDH is utilized to derive the pair-wise secret **K** for a ECU and SECU. Hence, we applied the execution time of two ECCDH algorithms from hardware experiment, namely FourQ on STM32 and NIST K-283 on W806, and then we had the result. With the help of a state-of-art high performance elliptic curve FourQ. In the best case, it took less than 31 ms to derive all pair-wise secret. In the worst case, it took less than 535 ms to finish. Regardless the fact that the recommended curve K-283 from standard NIST 800-56A on W806 leads to unacceptable result, in the best case more than 14 seconds and not less than 255 seconds in the worst case. We believe that our proposed scheme is still practical while using the high-end MCU and high performance cryptography algorithm.

In figure 9, the result of execution time for phase 3 group shared secret generation to derive group shared secret **SK** for ECU in the group. Here, we had the best result of both under 4 ms and the worst result that both took no longer than 62 ms to finish. In figure 10, we present the execution time for phase 4 session key generation to derive session key **SSK** for all ECU in a group. It is shown that in the best case, this procedure can be done under 3 ms for both MCUs, and in the worst case it took no more than 46 ms to complete for both. We argue that based on our simulation result that in STM32 it took total time of 614 ms to finish all three phases for the worst case of 35 ECUs in a group, it is practical that our proposed scheme can be applied in real environment.

TABLE VII
COMPARISON OF OUR WORK AND OTHERS

| Work | Message Complexity | Computation Complexity($N$=2) |
|---|---|---|
| OURS | $2N + 1$ | $5T_{sc} + 4T_h + 6T_{hkdf} + 4T_{hmac} + 2T_e \approx \mathbf{20.1}$ ms |
| Carvajal-Roca et al.[12] | $5N + 1$ | $6T_{sc} + 2T_h + 4T_e \approx \mathbf{21.7}$ ms |
| Musuroi et al.[10] | $4(2N - 1)$ | $T_s + 2T_v + 4T_{sc} + 4T_e \approx \mathbf{38.2}$ ms |

In order to compare our proposed scheme to others. We measured the total quantity of message being transmitted on the bus network after finishing the scheme. Here, we chose two similar works for comparison. Musuroi et al. proposed a scheme named full key exchange tree with SoECU, in order to derive group shared secret among ECU using a tree structure with a central node. Carvajal-Roca et al. proposed a key exchange scheme using identity-based encryption(IBE) and identity-based broadcast encryption(IBBE), also required a central security ECU. From Figure 11, it is shown that in the aspect of total message transmission, in the worst case, our work is 40.34% of Carvajal-Roca's work and 25.72% of Musuroi's work.

Table VII shows the message complexity and computation complexity of the three key management schemes, where $N$ stands for the number of ECU in a group, $T_s$ stands for the time consumption of digital signature generation, $T_v$ stands for the time consumption of digital signature verification, $T_{sc}$ stands for the time consumption of public key encryption, $T_h$ stands for the time consumption of hash function, $T_e$ stands for the time consumption of symmetric encryption/decryption, $T_{hkdf}$ stands for the time consumption of hash based key derivation function, $T_{hmac}$ stands for the time consumption of hash based message authentication code generation. All time consumption are measured on STM32, the time consumption of digital signature generation and verification are not shown in the hardware experiment above because they are not used in our scheme. Their result are, $T_s = 2.87$ ms and $T_v = 4.46$ ms. Based on figure 11 and table VII, we believe that our work has the advantage of lower communication and computation overhead compared to similar works.

## VI. CONCLUSION

In this work, we first summarized the researches on in-vehicle network vulnerability and countermeasures, then presented our novel efficient key management scheme using KEM based on ECCDH as well as other cryptography algorithm. To show that our proposed scheme is secure, we presented a formal analysis on the chosen-ciphertext attack for our KEM and an analysis for security property, namely cryptoperiod and security strength, based on NIST 800-57. Next, we presented two experiments to evaluate the performance of our proposed scheme. One is the hardware-based experiment, the other one is the software-based simulation. All together, we shown that for one our proposed protocol may be practical for reality use, and our work is superior in the aspects of communication and computation overhead compared to similar works.

In the future, we look forward to work on improving the time consumption of our KEM, which is using asymmetric cryptography algorithms. Inspired by [31], a possible speed-up might be implemented with the help of the concept of multi-exponentiations.

REFERENCES

[1] Autonomous Cars Market Size, Share, Growth & Forecast. [Online]. Available: https://www.fortunebusinessinsights.com/industry-reports/autonomous-cars-market-100141
[2] Robotaxi startup Pony.ai gains taxi license in China city. [Online]. Available: https://www.reuters.com/technology/robotaxi-startup-ponyai-gains-taxi-license-china-city-2022-04-24/
[3] New Car Hacking Research: 2017, Remote Attack Tesla Motors Again. [Online]. Available: https://keenlab.tencent.com/en/2017/07/27/New-Car-Hacking-Research-2017-Remote-Attack-Tesla-Motors-Again/
[4] Remote Exploitation of an Unaltered Passenger Vehicle. [Online]. Available: https://illmatics.com/Remote%20Car%20Hacking.pdf
[5] The Jeep Hackers Are Back to Prove Car Hacking Can Get Much Worse. [Online]. Available: https://www.wired.com/2016/08/jeep-hackers-return-high-speed-steering-acceleration-hacks/
[6] H. Sun, M. Chen, J. Weng, Z. Liu and G. Geng, "Anomaly Detection for In-Vehicle Network Using CNN-LSTM With Attention Mechanism," in IEEE Transactions on Vehicular Technology, vol. 70, no. 10, pp. 10880-10893, Oct. 2021, doi: 10.1109/TVT.2021.3106940.
[7] M. L. Han, B. I. Kwak and H. K. Kim, "Event-Triggered Interval-Based Anomaly Detection and Attack Identification Methods for an In-Vehicle Network," in IEEE Transactions on Information Forensics and Security, vol. 16, pp. 2941-2956, 2021, doi: 10.1109/TIFS.2021.3069171.
[8] X. Duan, H. Yan, D. Tian, J. Zhou, J. Su and W. Hao, "In-Vehicle CAN Bus Tampering Attacks Detection for Connected and Autonomous Vehicles Using an Improved Isolation Forest Method," in IEEE Transactions on Intelligent Transportation Systems, doi: 10.1109/TITS.2021.3128634.
[9] Y. Xun, Y. Zhao and J. Liu, "VehicleEIDS: A Novel External Intrusion Detection System Based on Vehicle Voltage Signals," in IEEE Internet of Things Journal, vol. 9, no. 3, pp. 2124-2133, 1 Feb.1, 2022, doi: 10.1109/JIOT.2021.3090397.
[10] A. Musuroi, B. Groza, L. Popa, and P.-S. Murvay, "Fast and Efficient Group Key Exchange in Controller Area Networks (CAN)," IEEE Trans. Veh. Technol., pp. 1–1, 2021, doi: 10.1109/TVT.2021.3098546.
[11] H. Li, V. Kumar, J.-M. Park, and Y. Yang, "Cumulative Message Authentication Codes for Resource-Constrained IoT Networks," IEEE Internet Things J., vol. 8, no. 15, pp. 11847–11859, Aug. 2021, doi: 10.1109/JIOT.2021.3074054.
[12] I. E. Carvajal-Roca, J. Wang, J. Du, and S. Wei, "A Semi-centralized Dynamic Key Management Framework for In-vehicle Networks," IEEE Trans. Veh. Technol., pp. 1–1, 2021, doi: 10.1109/TVT.2021.3106665.
[13] T.-Y. Youn, Y. Lee, and S. Woo, "Practical Sender Authentication Scheme for In-Vehicle CAN With Efficient Key Management," IEEE Access, vol. 8, pp. 86836–86849, 2020, doi: 10.1109/ACCESS.2020.2992112.
[14] B. Palaniswamy, S. Camtepe, E. Foo, and J. Pieprzyk, "An Efficient Authentication Scheme for Intra-Vehicular Controller Area Network," IEEE Trans.Inform.Forensic Secur., vol. 15, pp. 3107–3122, 2020, doi: 10.1109/TIFS.2020.2983285.
[15] H. Mun, K. Han, and D. H. Lee, "Ensuring Safety and Security in CAN-Based Automotive Embedded Systems: A Combination of Design Optimization and Secure Communication," IEEE Trans. Veh. Technol., vol. 69, no. 7, pp. 7078–7091, Jul. 2020, doi: 10.1109/TVT.2020.2989808.
[16] H. J. Jo, J. H. Kim, H.-Y. Choi, W. Choi, D. H. Lee, and I. Lee, "MAuth-CAN: Masquerade-Attack-Proof Authentication for In-Vehicle Networks," IEEE Trans. Veh. Technol., vol. 69, no. 2, pp. 2204–2218, Feb. 2020, doi: 10.1109/TVT.2019.2961765.

[17] M. Han, A. Wan, F. Zhang, and S. Ma, "An Attribute-Isolated Secure Communication Architecture for Intelligent Connected Vehicles," IEEE Trans. Intell. Veh., vol. 5, no. 4, pp. 545–555, Dec. 2020, doi: 10.1109/TIV.2020.3027717.

[18] B. Groza, L. Popa, and P.-S. Murvay, "Highly Efficient Authentication for CAN by Identifier Reallocation With Ordered CMACs," IEEE Trans. Veh. Technol., vol. 69, no. 6, pp. 6129–6140, Jun. 2020, doi: 10.1109/TVT.2020.2990954.

[19] B. Groza, S. Murvay, A. V. Herrewege, and I. Verbauwhede, "LiBrA-CAN: Lightweight Broadcast Authentication for Controller Area Networks," ACM Trans. Embed. Comput. Syst., vol. 16, no. 3, pp. 1–28, Jul. 2017, doi: 10.1145/3056506.

[20] S. Woo, H. J. Jo, I. S. Kim, and D. H. Lee, "A Practical Security Architecture for In-Vehicle CAN-FD," IEEE Trans. Intell. Transport. Syst., vol. 17, no. 8, pp. 2248–2261, Aug. 2016, doi: 10.1109/TITS.2016.2519464.

[21] S. Woo, H. J. Jo, and D. H. Lee, "A Practical Wireless Attack on the Connected Car and Security Protocol for In-Vehicle CAN," IEEE Trans. Intell. Transport. Syst., pp. 1–14, 2014, doi: 10.1109/TITS.2014.2351612.

[22] T. Sakon and Y. Nakamoto, "Poster: Simple key management scheme for in-vehicle system," in 2016 IEEE Vehicular Networking Conference (VNC), Columbus, OH, USA, Dec. 2016, pp. 1–2. doi: 10.1109/VNC.2016.7835976.

[23] T. Lenard, R. Bolboaca, and B. Genge, "LOKI: A Lightweight Cryptographic Key Distribution Protocol for Controller Area Networks," in 2020 IEEE 16th International Conference on Intelligent Computer Communication and Processing (ICCP), Cluj-Napoca, Romania, Sep. 2020, pp. 513–519. doi: 10.1109/ICCP51029.2020.9266192.

[24] T. Sugashima, D. K. Oka, and C. Vuillaume, "Approaches for Secure and Efficient In-Vehicle Key Management," DENSO TECHNICAL REVIEW Vol.21 2016 p. 10.

[25] E. Barker, "Recommendation for key management:: part 1 - general," National Institute of Standards and Technology, Gaithersburg, MD, NIST SP 800-57pt1r5, May 2020. doi: 10.6028/NIST.SP.800-57pt1r5.

[26] E. Barker, L. Chen, A. Roginsky, A. Vassilev, and R. Davis, "Recommendation for pair-wise key-establishment schemes using discrete logarithm cryptography," National Institute of Standards and Technology, Gaithersburg, MD, NIST SP 800-56Ar3, Apr. 2018. doi: 10.6028/NIST.SP.800-56Ar3.

[27] U. E. Larson, D. K. Nilsson and E. Jonsson, "An approach to specification-based attack detection for in-vehicle networks," 2008 IEEE Intelligent Vehicles Symposium, 2008, pp. 220-225, doi: 10.1109/IVS.2008.4621263.

[28] Information Technology Laboratory, "Digital Signature Standard (DSS)," National Institute of Standards and Technology, NIST FIPS 186-4, Jul. 2013. doi: 10.6028/NIST.FIPS.186-4.

[29] P. Mundhenk, A. Mrowca, S. Steinhorst, M. Lukasiewycz, S. A. Fahmy, and S. Chakraborty, "Open source model and simulator for real-time performance analysis of automotive network security," SIGBED Rev., vol. 13, no. 3, pp. 8–13, Aug. 2016, doi: 10.1145/2983185.2983186.

[30] P. Mundhenk et al., "Security in Automotive Networks: Lightweight Authentication and Authorization," ACM Trans. Des. Autom. Electron. Syst., vol. 22, no. 2, pp. 1–27, Mar. 2017, doi: 10.1145/2960407.

[31] E. Kiltz, "Chosen-Ciphertext Secure Key-Encapsulation Based on Gap Hashed Diffie-Hellman," in Public Key Cryptography – PKC 2007, vol. 4450, T. Okamoto and X. Wang, Eds. Berlin, Heidelberg: Springer Berlin Heidelberg, 2007, pp. 282–297. doi: 10.1007/978-3-540-71677-8_19.

[32] Arduino UNO R3. [Online]. Available: https://docs.arduino.cc/hardware/uno-rev3/

[33] WinnerMicro MCU W806 [Online]. Available: https://github.com/IOsetting/wm-sdk-w806

[34] M. L. Chavez, C. H. Rosete and F. R. Henriquez, "Achieving confidentiality security service for CAN," 15th International Conference on Electronics, Communications and Computers (CONIELECOMP'05), 2005, pp. 166-170, doi: 10.1109/CONIEL.2005.13.

[35] A. Mosenia and N. K. Jha, "A Comprehensive Study of Security of Internet-of-Things," IEEE Trans. Emerg. Topics Comput., vol. 5, no. 4, pp. 586–602, Oct. 2017, doi: 10.1109/TETC.2016.2606384.

[36] B. Groza and P.-S. Murvay, "Security Solutions for the Controller Area Network: Bringing Authentication to In-Vehicle Networks," IEEE Veh. Technol. Mag., vol. 13, no. 1, pp. 40–47, Mar. 2018, doi: 10.1109/MVT.2017.2736344.

[37] E. Aliwa, O. Rana, C. Perera, and P. Burnap, "Cyberattacks and Countermeasures for In-Vehicle Networks," ACM Comput. Surv., vol. 54, no. 1, pp. 1–37, Apr. 2021, doi: 10.1145/3431233.

[38] P. Thirumavalavasethurayar and T. Ravi, "Implementation of Replay Attack in Controller Area Network Bus using Universal Verification Methodology," in 2021 International Conference on Artificial Intelligence and Smart Systems (ICAIS), Coimbatore, India, Mar. 2021, pp. 1142–1146. doi: 10.1109/ICAIS50930.2021.9395871.

[39] E. Aliwa, O. Rana, C. Perera, and P. Burnap, "Cyberattacks and Countermeasures for In-Vehicle Networks," ACM Comput. Surv., vol. 54, no. 1, pp. 1–37, Apr. 2021, doi: 10.1145/3431233.

[40] P.-S. Murvay and B. Groza, "Security Shortcomings and Countermeasures for the SAE J1939 Commercial Vehicle Bus Protocol," IEEE Trans. Veh. Technol., vol. 67, no. 5, pp. 4325–4339, May 2018, doi: 10.1109/TVT.2018.2795384.

[41] B. Groza and S. Murvay, "Efficient Protocols for Secure Broadcast in Controller Area Networks," IEEE Trans. Ind. Inf., vol. 9, no. 4, pp. 2034–2042, Nov. 2013, doi: 10.1109/TII.2013.2239301.

[42] A. Perrig, D. Song, R. Canetti, J. D., and B. Briscoe, "Timed Efficient Stream Loss-Tolerant Authentication (TESLA): Multicast Source Authentication Transform Introduction," RFC Editor, RFC4082, Jun. 2005. doi: 10.17487/rfc4082.

[43] VECTOR CANoe. [Online]. Available: https://www.vector.com/int/en/products/products-a-z/software/canoe/

[44] V. Shoup, "Using Hash Functions as a Hedge against Chosen Ciphertext Attack," in Advances in Cryptology — EUROCRYPT 2000, vol. 1807, B. Preneel, Ed. Berlin, Heidelberg: Springer Berlin Heidelberg, 2000, pp. 275–288. doi: 10.1007/3-540-45539-6_19.

[45] Bosch CAN Specification 2.0. [Online]. Available: http://esd.cs.ucr.edu/webres/can20.pdf

[46] Automotive ECU: Core Component for Connected Cars. [Online]. Available: https://www.syrma.com/ecu/

[47] X. Ying, G. Bernieri, M. Conti, and R. Poovendran, "TACAN: transmitter authentication through covert channels in controller area networks," in Proceedings of the 10th ACM/IEEE International Conference on Cyber-Physical Systems, Montreal Quebec Canada, Apr. 2019, pp. 23–34. doi: 10.1145/3302509.3313783.

[48] D. A. Fisher, J. M. McCune, and A. D. Andrews, "Trust and Trusted Computing Platforms:," Defense Technical Information Center, Fort Belvoir, VA, Jan. 2011. doi: 10.21236/ADA536188..

[49] D. Eastlake and T. Hansen, "US Secure Hash Algorithms (SHA and SHA-based HMAC and HKDF)," RFC Editor, RFC6234, May 2011. doi: 10.17487/rfc6234.

[50] TCG TPM 2.0 Automotive Thin Profile For TPM Family 2.0; Level 0 Specification Version 1.01 Revision 14 January 18, 2018. [Online]. Available: https://trustedcomputinggroup.org/wp-content/uploads/TCG_TPM_2.0_Automotive-Thin_Profile_v1.1-r14_Public-Review.pdf

[51] R. Cramer and V. Shoup, "Design and Analysis of Practical Public-Key Encryption Schemes Secure against Adaptive Chosen Ciphertext Attack," SIAM J. Comput., vol. 33, no. 1, pp. 167–226, Jan. 2003, doi: 10.1137/S0097539702403773.

[52] C. Costello and P. Longa, "FourQ: four-dimensional decompositions on a Q-curve over the Mersenne prime." Advances in Cryptology - ASIACRYPT 2015. Available: https://eprint.iacr.org/2015/565

[53] R. Canetti, V. Kolesnikov, C. Rackoff, and Y. Vahlis, "Secure Key Exchange and Sessions without Credentials," in Security and Cryptography for Networks, vol. 8642, M. Abdalla and R. De Prisco, Eds. Cham: Springer International Publishing, 2014, pp. 40–56. doi: 10.1007/978-3-319-10879-7_3.

[54] STM32F407. [Online]. Available: https://www.st.com/en/microcontrollers-microprocessors/stm32f407-417.html#overview

[55] Tiny ECDH / ECC in C. [Online]. Available: https://github.com/kokke/tiny-ECDH-c
[56] High performance implementation of elliptic curve 25519. [Online]. Available: https://github.com/msotoodeh/curve25519
[57] FourQlib v3.0 (C Edition). [Online]. Available: https://github.com/microsoft/FourQlib/tree/master/FourQ_ARM
[58] Tiny AES. [Online]. Available: https://github.com/kokke/tiny-AES-c
[59] A. Happel, E. Sparrer, and P. Deck, "CAN FD: Measuring and reprogramming," CAN Newsletter 3/2014 p. 4. Available: https://cdn.vector.com/cms/content/know-how/_technical-articles/CAN_FD_Measurement_CiA_PressArticle_201409_EN.pdf
[60] R. De Andrade, K. N. Hodel, J. F. Justo, A. M. Lagana, M. M. Santos, and Z. Gu, "Analytical and Experimental Performance Evaluations of CAN-FD Bus," IEEE Access, vol. 6, pp. 21287–21295, 2018, doi: 10.1109/ACCESS.2018.2826522.